*Introduction.* — In a model-theoretic context of long-range correlations in DNA, Roche *et al*. [5] have studied Fibonacci polyGC quasiperiodic sequences based on the inflation rule G→GC and C→G, which gives successively G, GC, GCG, GCGGC, GCGGCGCG, . . . for sequences of length 1, 2, 3, 5, 8, . . . respectively. It appears that such Fibonacci sequences can be used as prototypes for strongly correlated DNA segments as long as ~160 nm (*i.e*., ~450 bp).

More than fifty years ago, Weyl [9] showed that the symmetry in biological phenotypes can be quantified by basic group theory, as shown for example in Fig. 1. That Fibonacci and other symmetries may exist in the sequences of DNA's, RNA's and certain proteins is suggested by the recent findings of motif pattern sequences—— the so-called *seqlets* [3, 4] that occur repeatedly in proteins. Typically featuring about 10 amino acid residue positions that are fixed to within conservative substitutions but usually separated by a number of prescribed gaps with arbitrary residue content, the existence of protein-fragment seqlets poses an intriguing question: Do certain seqlets have a sequence with a symmetry precisely quantifiable by group theory? This question is answered affirmatively in the present communication. We have found a precise modular Fibonacci sequence in a continuous gap-free 10-residue seqlet with either 3 or 4 conservative amino acid substitutions. Unambiguous statistical significance is attached to this modular Fibonacci seqlet, for it occurs nine times in the SWISS-Prot/TrEMBL database [1,2] of natural proteins. As in many cases of phenotypic symmetry [9], cyclic group theory here again provides the medium for the analysis.

*Nucleotide and codon representative cyclic groups.* — We start with a purely physical $C_4$ (cyclic group of order 4) representation for the nucleotides. As shown in Table 1, the number of oxygen and nitrogen atoms $n_O$ and $n_N$ along with the molecular ring numbers r engender the complex number $z = (i)^{n_O} (-1)^{n_N - r}$ for each nucleotide. Constituting a representation of $C_4$, the four complex numbers have the form $z = \exp(i n \pi / 2)$ where [6,7] $n(U) = 0$, $n(C) = 1$, $n(A) = 2$ and $n(G) = 3$. Note that under the complementarity transpositions $U \leftrightarrow A$, $C \leftrightarrow G$, we have $z \rightarrow -z$ for all z.

Next, we make the biologically appropriate foliation of the $C_4$ cyclic group for the nucleotides into a $C_{64}$ (cyclic group of order 64) representation for the codons. Let $5'(\mathcal{N}_- \mathcal{N} \mathcal{N}_+) 3'$ denote a generic codon with the successive triplet of nucleotides $\mathcal{N}_-$, $\mathcal{N}$, $\mathcal{N}_+$. Then the codon's $C_{64}$ representative is defined by

$$Z = (z_-)^{1/4} z (z_+)^{1/16} = \exp(iN\pi/32) \qquad (1)$$

in which $z_-$, $z$, $z_+$ are the $C_4$ representatives of the respective nucleotides and the principal-branch roots are taken by the fractional exponents on $z_-$ and $z_+$ in the middle member of (1). The codon number *N* introduced by implicit definition in the final member of (1) is therefore given by [6,7]

$$N = 4 n_- + 16 n + n_+ \qquad (2)$$

Thus, for example, with $5'(\mathcal{N}_- \mathcal{N} \mathcal{N}_+) 3' = $ AUG for Methionine, we have $N = 4n(A) + 16n(U) + n(G) = 11$. Observe that *N* runs from 0 to 63 along with the codons in the natural progression of the universal genetic code (see Table 2). Finally, the code itself assigns one of the 20 amino acids or the "stop" signal to each codon, and thus concomitantly to its codon number *N*. Hence, a value for *N* implies an amino acid or the "stop" signal (the latter, in the cases $N = 34$, 35 or 50). Moreover, a sequence of *Z*'s of the form (1), or equivalently a sequence of *N* values, implies a specific polypeptide which may or may not be a continuous gap-free seqlet.

*Modular Fibonacci sequences.* — Now let $Z_{k-1}$, $Z_k$, $Z_{k+1}$ denote the $C_{64}$ representatives of three successive codons in an mRNA. Suppose that through the formation mechanism of the mRNA the representative $Z_{k+1}$ depends on its two predecessors $Z_k$, $Z_{k-1}$ in a symmetrical-function manner over a range of integer *k* values, thereby representing a group-theoretic direct-product surjection of the form $C_{64}(k) \times C_{64}(k-1) \to C_{64}(k+1)$. The only uniformly consistent way that this can happen is for the $C_{64}$ representatives to satisfy $Z_{k+1} = Z_k Z_{k-1}$ for all *k* in the range, a relation equivalent to the modular Fibonacci sequence condition

$$N_{k+1} = (N_k + N_{k-1}) \pmod{64} \tag{3}$$

Here "mod 64" instructs one to "subtract 64 from the sum if it is greater than 63", for *Z* in (1) is unchanged by $N \to (N - 64)$ and *N* ranges from 0 to 63 by definition in (2).

A sequence of $N_k$'s (to which there is an associated polypeptide) follows from (3) if one prescribes $N_1$, $N_2$ and the range of *k*. As a prime example, consider the biological sequence that features $N_1 = 11$ for **M** (see Table 2), the "start" signal for a protein, and $N_2 = 33$, which is **M**'s partner under nucleotide complementarity: $N_1 = 11 \Leftrightarrow \text{AUG} \leftrightarrow \text{UAC} \Leftrightarrow N_2 = 33$. Then (3) employed recursively produces the sequence

11  33  44  13  57  6  63  5  4  9  13  22  35

Note that we have terminated the sequence at $k = 13$, because $N_{13} = 35$ signals "stop". By using Table 2 the sequence translates into the polypeptide

**M Y D V S L G L L I V P** (stop)

which finally yields the modular Fibonacci seqlet

**Y D V S L G L L I V** (4)

In (4) we have deleted the **M** (interpreted as a "start" signal) and the **P** (for terminal deletion symmetry). The standard equivalence classes for amino acid conservative substitutions are {**F, Y**}, {**L, I, M, V**}, {**S, T**}, {**P**}, {**A, G**}, {**H**}, {**Q, N**}, {**K. R**}, {**D. E**}, {**C**}, [**W**], as permitted by the SWISS-Prot/TrEMBL BLAST program [1,2]. With (4) as the query, the SWISS-Prot/TrEMBL database (containing 137,811 non-redundant sequences with 50,431,180 total letters when accessed) identified the following nine protein segments as biological equivalents to (4):

```
    >gi 3122441 sp 021077 NU1M_MYXGL       NADH-ubiquinone oxidoreductase chain 1
         Length = 318     Identities = 7/10 (70%), Positives = 10/10 (100%)
Query:      1      YDVSLGLLIV            10
                   Y+V+LGL+IV
Subject:   144     YEVTLGLMIV           153

    >gi 266658 sp P29920 NQO8_PARDE        NADH-quinone oxidoreductase chain 8
           (NADH dehydrogenase: NDH-1, chain 8)
         Length = 345     Identities = 7/10 (70%), Positives = 10/10 (100%)
Query:      1      YDVSLGLLIV            10
                   Y+VSLGL+I+
Subject:   157     YEVSLGLIII           166

    >gi 1171804 sp P42032 NUOH_RHOCA       NADH-quinone oxidoreductase chain H
           (NADH dehydrogenase: NDH-1, chain H)
         Length = 345     Identities = 7/10 (70%), Positives = 10/10 (100%)
Query:      1      YDVSLGLLIV            10
                   Y+VS+GL+IV
Subject:   157     YEVSMGLIIV           166

    >gi 6647674 sp Q9ZCF7 NUOH_RICPR       NADH-quinone oxidoreductase chain H
           (NADH dehydrogenase: NDH-1, chain H)
         Length = 339     Identities = 6/10 (60%), Positives = 10/10 (100%)
Query:      1      YDVSLGLLIV            10
                   Y+VS+GL+I+
Subject:   157     YEVSMGLVII           166

    >gi 1352538 sp P48899 NUIM_CYACA       NADH- ubiquinone oxidoreductase chain 1
         Length = 344     Identities = 6/10 (60%), Positives = 10/10 (100%)
Query:      1      YDVSLGLLIV            10
```

```
                    Y+VS+GL+I+
Subject:    165     YEVSIGLIII            174

   >gi 14194971 sp O79546 NUIM_DINSE         NADH- ubiquinone oxidoreductase chain 1
         Length = 321     Identities = 6/10 (60%), Positives = 10/10 (100%)
Query:      1       YDVSLGLLIV            10
                    Y+V+LGL+I+
Subject:    148     YEVTLGLIII            157

   >gi 2499233 sp Q96182 NUIM_POLOR          NADH- ubiquinone oxidoreductase chain 1
         Length = 319     Identities = 6/10 (60%), Positives = 10/10 (100%)
Query:      1       YDVSLGLLIV            10
                    Y+V+LGL+I+
Subject:    146     YEVTLGLIII            155

   >gi 462756 sp P26845 NUIM_MARPO           NADH-ubiquinone oxidoreductase chain 1
         Length = 328     Identities = 6/10 (60%), Positives = 10/10 (100%)
Query:      1       YDVSLGLLIV            10
                    Y+VS+GLL++
Subject:    150     YEVSIGLLLI            159

   >gi 3122451 sp Q37556 NUIM_METSE          NADH-ubiquinone oxidoreductase chain 1
         Length = 334     Identities = 6/10 (60%), Positives = 10/10 (100%)
Query:      1       YDVSLGLLIV            10
                    Y+VS+GL+I+
Subject:    157     YEVSIGLIII            166
```

*Statistical significance. —* Observe that (3) prescribes the 9 amino acids that follow **Y** in (4), with $N_1 = 11$ and $N_2 = 33$. For an incidentally random sequence, the probability of a single appearance of a biological equivalent of (4) with **Y** in the first position and j conservative substitutes in the subsequent nine positions is given approximately by $\phi_j = (9!/(9-j)!\, j!)$ $(.05)^{10-j} (.11)^j (5.04 \times 10^7)$, in which (.05) is the average probability for any one of the 20 amino acids, (.11) is the average probability for an amino acid or one of its biological equivalents, and $(5.04 \times 10^7)$ is the approximate number of total letters, *i.e.*, chances for a hit in the database; for j = 3 and 4 we therefore have $\phi_3 = 4.40 \times 10^{-3}$ and $\phi_4 = 1.45 \times 10^{-2}$. Thus, even with suitable latitude given to evolutionary correlations between the SWISS-Prot/TrEMBL protein fragments cited above, it follows that the incidental random-sequence composite probability for three j = 3 and six j = 4 conformances is small enough to preclude an incidental origin for the accord with virtual certainty. Therefore, the modular Fibonacci seqlet shown in (4) can be deemed to be genuinely biophysical.

*A possible dynamical origin for modular Fibonacci seqlets.* — In conclusion, it should be emphasized that the nucleotide group $C_4$, the codon group $C_{64}$, and hence a modular Fibonacci sequence in proteins have a precise biophysical underpinning in terms of nucleotide and codon aerobic-atom content and molecular ring structure, as expressed by and $n_O$, $n_N$ and r. The z's of $C_4$ and the Z's of $C_{64}$ may perhaps also have a quantum physics origin as wavefunction prefactors that result from the nucleotide and codon production reactions. Consonant with their aerobic-atom content and molecular ring structure, the z's may have originated dynamically as relative prefactors in the nucleotide groundstate wavefunctions; if so, the phase angles in the z, $n\pi/2$, may be interpreted as residual Berry phase angles [8] from the nucleotide production process. Furthermore, if the $Z_k$'s are residual Berry phase factors from the codon production process, then the relation $Z_{k+1} = Z_k Z_{k-1}$ and its equivalent (3) minimize an energy of the form $-\varepsilon \, \text{Re}(\sum_k Z^*_{k+1} \, Z_k \, Z_{k-1})$, where the constant $\varepsilon$ is the magnitude of an intercodon binding energy increment in the mRNA groundstate. Irrespective of the underlying mechanism, however, the biophysical genuineness of the modular Fibonacci sequence is clearly indicated statistically.

*Acknowledgements.* —   M.D. is grateful for the helpful communications received from Elisabeth Gasteiger on the BLAST program procedures.  G. R. wishes to thank Robert Kuniewicz and Neville Kallenbach for their suggestions regarding the SWISS-Prot/TrEMBL database.  The photograph of the *Nautilus* was kindly taken by Kevin Aires (courtesy of K. R. Design).

MCS: 92D20, 62P10

Figure 1

Half-shell of a pearly *Nautilus* in which the edge of the coil is a near-perfect logarithmic spiral [5], a curve such that the rates of rotation and dilatation are proportional [hence yielding $\theta \propto \ln(r/r_o)$]. The self-similar air chambers increase in diameter according to a fractionally-muted Fibonacci relation [*viz.*, $d_k = (\text{const})(f_k)^{1/8}$ for the $k^{th}$ chamber, where $f_k = 1, 2, 3, 5, 8, 13, 21, \ldots$ is the primary Fibonacci series].

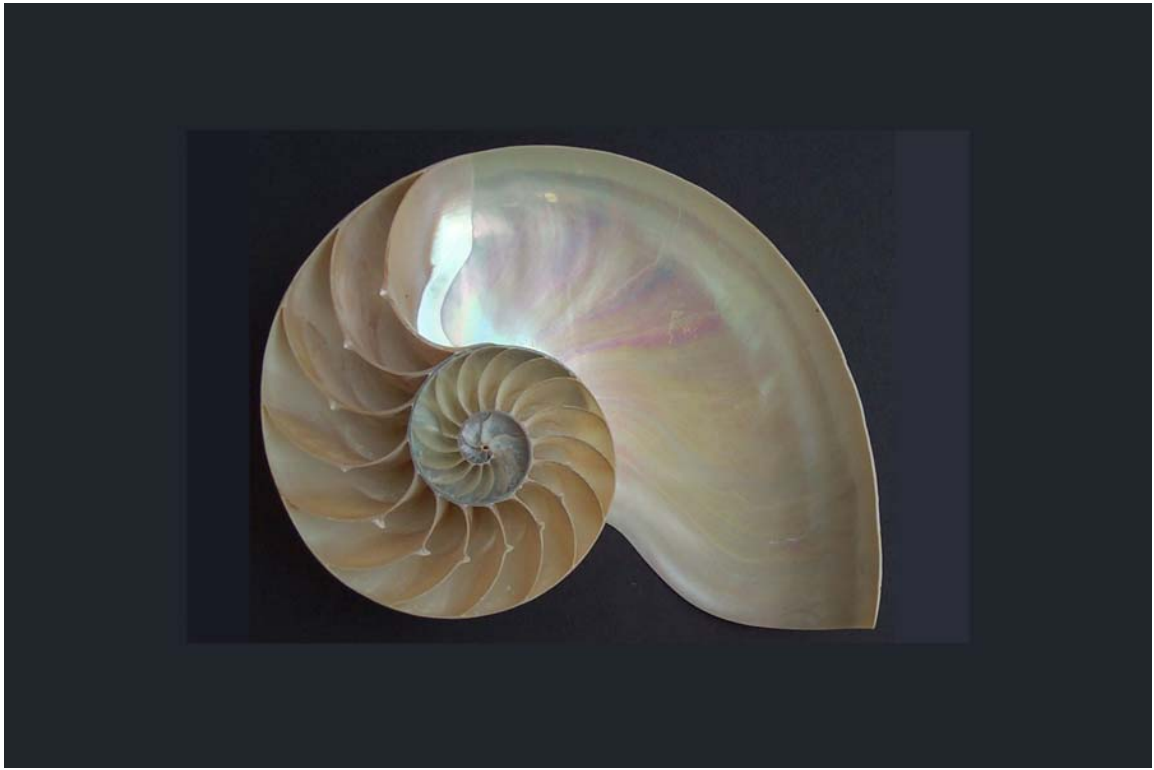

Table 1

The nucleotide $C_4$ representatives $z = (i)^{n_O}(-1)^{n_N - r}$. Here $n_O$, $n_N$, and r are the number of oxygen atoms, nitrogen atoms, and rings in the nucleotide molecule, the quantities that determine its z value.

|  |  | $n_O$ | $n_N$ | r | z |
|---|---|---|---|---|---|
|  | U, uracil | 2 | 2 | 1 | 1 |
| pyrimidines | T, thymine | 2 | 2 | 1 | 1 |
|  | C, cytosine | 1 | 3 | 1 | $i$ |
| purines | A, adenine | 0 | 5 | 2 | −1 |
|  | G, guanine | 1 | 5 | 2 | −$i$ |

Table 2

The universal genetic code annotated with codon numbers. The code associates 61 mRNA codons—ordered triplets of the nucleotides U, C, A, G—with the 20 lifeform amino acids. In particular, AUG for **M** also serves as a "start" signal for a protein, while UAA, UAG and UGA signal "stop" by not being associated with any amino acid. The codon number *N* [given by Eq. (2)] appears to the left of each respective codon, and the associated amino acid along with its letter symbol appears to the right of its codon correspondents.

| | | | | | | | | | | | |
|---|---|---|---|---|---|---|---|---|---|---|---|
| 0 | UUU | Phenylalanine | 16 | UCU | | 32 | UAU | Tyrosine | 48 | UGU | Cysteine |
| 1 | UUC | **F** | 17 | UCC | Serine | 33 | UAC | **Y** | 49 | UGC | **C** |
| 2 | UUA | | 18 | UCA | **S** | 34 | UAA | (stop) | 50 | UGA | (stop) |
| 3 | UUG | | 19 | UCG | | 35 | UAG | (stop) | 51 | UGG | Tryptophan **W** |
| 4 | CUU | Leucine | 20 | CCU | | 36 | CAU | Histidine | 52 | CGU | |
| 5 | CUC | **L** | 21 | CCC | Proline | 37 | CAC | **H** | 53 | CGC | Arginine |
| 6 | CUA | | 22 | CCA | **P** | 38 | CAA | Glutamine | 54 | CGA | **R** |
| 7 | CUG | | 23 | CCG | | 39 | CAG | **Q** | 55 | CGG | |
| 8 | AUU | | 24 | ACU | | 40 | AAU | Asparagine | 56 | AGU | Serine |
| 9 | AUC | Isoleucine | 25 | ACC | Threonine | 41 | AAC | **N** | 57 | AGC | **S** |
| 10 | AUA | **I** | 26 | ACA | **T** | 42 | AAA | Lysine | 58 | AGA | Arginine |
| 11 | AUG | Methionine **M** (start) | 27 | ACG | | 43 | AAG | **K** | 59 | AGG | **R** |
| 12 | GUU | | 28 | GCU | | 44 | GAU | Aspartic acid | 60 | GGU | |
| 13 | GUC | Valine | 29 | GCC | Alanine | 45 | GAC | **D** | 61 | GGC | Glycine |
| 14 | GUA | **V** | 30 | GCA | **A** | 46 | GAA | Glutamic acid | 62 | GGA | **G** |
| 15 | GUG | | 31 | GCG | | 47 | GAG | **E** | 63 | GGG | |